\documentclass[twocolumn,aps,showpacs]{revtex4}
\usepackage{epsf,graphicx}
\usepackage{subfigure}
\usepackage[latin1]{inputenc}
\usepackage{amsmath}
\usepackage{amsfonts}
\usepackage{amssymb}

\newcommand {\pdif}[1]{\frac{\partial}{\partial #1}}

\newcommand {\VEC}[1]{\textbf{#1}}
\newcommand {\psixtau}{\Psi(\VEC{x},\tau)}
\newcommand {\psixtaust}{\Psi^*(\VEC{x},\tau)}

\begin{document}

\title{Systematic Semiclassical Expansion\\ for Harmonically
Trapped Ideal Bose Gases}

\author{B. Kl\"under${}^1$ and A. Pelster${}^2$}
\affiliation{${}^1$Arnold Sommerfeld Center for Theoretical Physics, Fakult\"at f\"ur Physik, 
Ludwig-Maximilians-Universit\"at M\"unchen, Theresienstrasse 37, 80333 M\"unchen, Germany\\
${}^2$Universit\"at Duisburg-Essen, Fachbereich Physik, Campus Duisburg, Lotharstrasse 1, 47057 Duisburg, Germany }

\date{\today}

\begin{abstract}
Using a field-theoretic  approach, we systematically
generalize the usual semiclassical approximation for a harmonically trapped ideal Bose gas in such a way
that its range of applicability is essentially extended. With this we
can analytically calculate  thermodynamic properties 
even for small particle numbers. In particular, it now becomes  possible
to determine the critical temperature as well as the temperature dependence
of both heat capacity and condensate fraction in low-dimensional
traps, where the standard semiclassical approximation is not even applicable.
\end{abstract}
\maketitle
%
\section{Introduction}
The field of ultracold Bose gases attains at present a lot of attention  due to an improved experimental accessibility 
within the last decade. Many different theoretical approaches are used to treat these trapped dilute quantum gases. 
Although isolated Bose gases should, in principle, be described within the micro-canonical ensemble, one commonly 
applies the technically  more efficient canonical or grand-canonical descriptions 
\cite{Glaum,Grossmann,Ketterle,Cornell,Yukalov,Wilkens,Holthaus1,Scully1,Holthaus2,Holthaus3,Weiss,Politzer,Kirsten,Yukalova,Gajda,Pethick,Holthaus4,Holthaus5,Scully2,Pitaevskii}. 
This is justified as experiments often use a large number of bosons.
It is a common belief in quantum statistics that, at least in the thermodynamic limit $N{\rightarrow}\infty$, 
all ensembles should converge to one and the same result. 
However, we note that some peculiar exceptions are known for particle counting statistics
as discussed, for instance, in Refs.~\cite{Glaum,ziff,navez}.
From a theoretical point of view, the grand-canonical ensemble has the advantage that it provides an analytical 
description, whereas the canonical approach is limited to numerical results for moderate particle numbers. 
As experiments with ultracold Bose gases are always realised with a finite number of particles,  the fundamental 
question arises how to study finite-size effects for the thermodynamic properties of trapped Bose gases most 
efficiently. \\
\indent To analyse this problem systematically, we introduce and compare two different approaches.
In Sect.~\ref{Numerically Exact Thermodynamics} we briefly rederive the well-known grand-canonical description without 
using the order parameter concept for a harmonically confined  ideal Bose gas. The thermodynamic properties can only  
be calculated numerically in this theory.
In Sect.~\ref{with order parameter} we introduce another grand-canonical description of the trapped Bose gas which is 
analytical as it relies on the order parameter concept. Introducing an order parameter is an essential approximation 
for finite systems and leads to different results for the respective thermodynamic quantities compared to the theory 
without order parameter. However these differences vanish in the thermodynamic limit and turn out to be negligibly 
small for experimentally realistic system sizes. 
Moreover, generalizing a formalism developed in Ref.~\cite[Appendix 7A]{Kleinert}, this approach 
extends the usual semiclassical approximation \cite{Grossmann,Ketterle,Pitaevskii,Pethick} to a systematic semiclassical 
expansion which yields yet unknown
analytical results for the thermodynamic quantities in the superfluid phase. In particular, we will calculate the 
critical temperature $T_c$ as well as the temperature dependence of the condensate fraction $N_0/N$ and the heat 
capacity $C_V$ for $D{=}1,2,3$ dimensions up to the order of the semiclassical expansion which was not accessible before.

\section{Approach Without Order Parameter}
\label{Numerically Exact Thermodynamics}
We start with briefly rederiving the well-known grand-canonical description of an ideal Bose gas.
The general expression for the grand-canonical potential of an ideal Bose gas is given by \cite{schwabl}
\begin{eqnarray}
	\Omega &=& \frac{1}{\beta}\sum_{\VEC{n}} 
		   \log\big[1-e^{-\beta(E_{\VEC{n}}-\mu)}\big]\label{freie.energie.allg}\,,
\end{eqnarray} 
where $\VEC{n}$ describes the one-particle quantum numbers.  Here $E_\VEC{n}$, $\beta{=}1/k_B T$, and $\mu$ denote the 
energy levels of the system, the inverse temperature, and the chemical potential, respectively.
We specify Eq.~(\ref{freie.energie.allg}) for the case of an ideal Bose gas which is trapped in an isotropic harmonic 
potential of the form $V(\VEC{x}){=}M\omega^2\VEC{x}^2/2$, where $M$ and $\omega$ denote the mass of a bosonic 
particle and the trap frequency, respectively. The  one-particle energy eigenvalues of a harmonic oscillator are  
$E_{\VEC{n}}=\hbar\omega(n_1+\ldots+n_D+D/2)$ in $D$ dimensions, so the grand-canonical 
potential (\ref{freie.energie.allg}) specifies to
\begin{equation}
	\Omega = -\frac{1}{\beta}\sum\limits_{k=1}^{\infty}
	       \frac{e^{\beta(\mu-E_0)k}}{k(1-e^{-\hbar\omega\beta
	        k})^D}\label{freie.energie.harm}\,.
\end{equation}
For numerical calculations it turns out to be useful to 
follow Ref.~\cite{GlaumPhD} and reexpress (\ref{freie.energie.harm})  by using the polylogarithms 
\begin{equation}
  \zeta_a(x)=\sum_{k=1}^\infty\frac{x^k}{k^a}\label{polylog}
\end{equation}
and the identity
\begin{equation}
\frac{1}{(1{-}z)^D}{=}\sum_{m=0}^{\infty}\binom{m{+}D{-}1}{m} z^m\,,
\end{equation}
so we obtain
\begin{eqnarray}
   \Omega 
    &=&   -\frac{1}{\beta}\sum_{m=0}^\infty\binom{m{+}D{-}1}{m}\,
          \zeta_1\left( e^{\beta(\mu-E_0 -m\hbar\omega)} \right)
	  \,.\qquad
\end{eqnarray}
With $N{=}{-}\partial\Omega/\partial\mu$ one gets for the particle number equation 
\begin{eqnarray}
   N  &=&  \sum_{m=0}^\infty \binom{m{+}D{-}1}{m}\,\zeta_0\left( e^{\beta(\mu-E_0 -m\hbar\omega)}\right)\,,
          \label{N.mit.geometrischer.reihe.harm}
\end{eqnarray}
which can be solved numerically for the chemical potential $\mu$ once the particle number $N$ is given.
In this approach we consider the resulting condensate fraction
\begin{equation}
\frac{ N_{0}}{ N}{=}\frac{1}{ N[e^{\beta(E_0{-}\mu)}{-}1]}\label{cond.fraction}
\end{equation}
as a function of the temperature and define the point where the curvature is maximal as
the critical temperature $ T_{c}$.
\begin{figure}[t]
\begin{center}
\begin{minipage}[b]{0.7\linewidth}
\setlength{\unitlength}{1\linewidth}
\begin{picture}(1,0.6)
\put(0,-0.03){\small$(\mu{-}E_0)/\hbar\omega$}
  \put(1.01,0.56){\small$ T/ T_c$}
\includegraphics[width=1\linewidth]{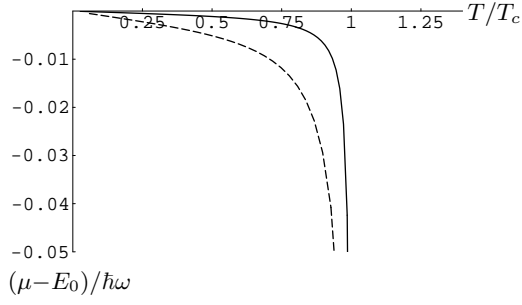}
\end{picture}
\end{minipage}
\end{center}
\caption{Dimensionless chemical potential $(\mu{-}E_0)/\hbar\omega$ versus $ T/ T_c$ in $D{=}3$ dimensions determined 
from (\ref{N.mit.geometrischer.reihe.harm}) for $ N{=}10^3$ particles (dashed line) and $ N{=}10^4$ (solid line) 
for  $\omega{=}2\pi{\cdot} 40~\mbox{Hz}$.  The critical temperatures are $ T_c{=}16.75~\mbox{nK}$ for $ N{=}10^3$ 
and $ T_c{=}37.56~\mbox{nK}$ for $ N{=}10^4$, respectively.}
\label{abb:chem.potential}
\end{figure}
Figure \ref{abb:chem.potential} shows the chemical potential $\mu$ in $D{=}3$ dimensions determined from 
(\ref{N.mit.geometrischer.reihe.harm}) as a function of temperature $T$ for a given particle number $N$.
One observes that $\mu$ remains smaller than the ground-state energy $E_0$ for all temperatures and approaches $E_0$ 
in the limit $T{\downarrow}0$. Moreover, one can see that $(\mu{-}E_0)/\hbar\omega$ gets smaller for $T{\leq}T_c$  
if the particle number $ N$ is increased.\\
The heat capacity $C_V{=}\partial U/\partial T\vert_{V,N}$ is derived from the internal 
energy $U{=}\Omega{+} T S{+} \mu N$ and yields
\begin{eqnarray}
   && C_V  {=}  k_B (\hbar\omega\beta)^2\label{Cv.oo.numerisch}\\
           &&\times\sum\limits_{m=0}^\infty m\binom{m{+}D{-}1}{m}\,
              \zeta_{-1}\left( e^{\beta(\mu-E_0 -m\hbar\omega)}\right)\nonumber\\
	&& \times\left[m{-}
	   \frac{\sum\limits_{m'=0}^\infty m'\binom{m'{+}D{-}1}{m'}\,
\zeta_{-1}\left( e^{\beta(\mu-E_0 -m'\hbar\omega)}\right)}{\sum\limits_{m''=0}^\infty \binom{m''{+}D{-}1}{m''}\,
\zeta_{-1}\left( e^{\beta(\mu-E_0 -m''\hbar\omega)}\right)}\right]\nonumber\,.
\end{eqnarray}
\begin{figure}[t]
\begin{center}
\begin{minipage}[b]{0.7\linewidth}
\setlength{\unitlength}{1\linewidth}
\begin{picture}(1,0.7)
\put(0,0.64){\small$ C_V/Nk_B$}
  \put(1,0.04){\small$ T/ T_c$}
\includegraphics[width=1\linewidth]{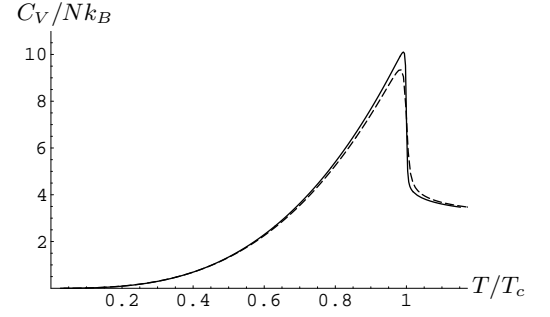}
\end{picture}
\end{minipage}
\end{center}
\caption{Heat capacity (\ref{Cv.oo.numerisch}) versus $ T/ T_c$ in $D{=}3$ dimensions for $ N{=}10^4$ (dashed line) 
and $ N{=}10^5$ (solid line) for $\omega{=}2\pi{\cdot}40~\mbox{Hz}$. The critical temperatures are 
$T_c{=}37.56~\mbox{nK}$ for $ N{=}10^4$ and $ T_c{=}82.46~\mbox{nK}$ for $ N{=}10^5$, respectively.}
\label{abb:cv-num}
\end{figure}

\noindent One can see in Fig.~\ref{abb:cv-num} that $ C_V$  in $D{=}3$ dimensions has its maximum at 
$T{\approx} T_c$ and vanishes exponentially fast in the limit $ T{\downarrow}0$. The heat capacity $C_V$ gets 
larger for $T{\leq} T_c$ and
 smaller for $ T{>} T_c$ if  the particle number $ N$ is increased. Moreover, one obtains that the slope at $ T{=} T_c$ 
increases as well for larger $ N$, but does not diverge.
With this we have shown exemplarily that phase transitions do not occur in finite systems.
However, $ \mu$ and $ C_V$ for fixed 
$T/ T_c$ seem to tend towards a limit for large particle numbers $ N$ and thus, we expect the emergence of a sharp phase 
transition in the thermodynamic limit $ N{\rightarrow}\infty$. In the next section we use this observation as a 
motivation to introduce an analytical approach for describing a trapped Bose gas with the help of an order parameter.
\section{Approach With Order Parameter}
\label{with order parameter}
We start with the functional integral approach to the grand-canonical partition function of a harmonically trapped ideal 
Bose gas \cite{Negele}
\begin{equation}
    Z=\oint\mathcal D\psi^*\mathcal D\psi\,e^{-\mathcal A[\psi^*,\psi]/\hbar}\,,\label{grosskan.Z.funtionalint.allg}
\end{equation}
where one integrates over all possible bosonic Schr\"odinger fields $\psi^*(\VEC{x},\tau),\psi(\VEC{x},\tau)$ which are 
periodic in imaginary time $\tau$ with period $\hbar \beta$. 
The Euclidean action $\mathcal A[\psi^*,\psi]$ reads
\begin{eqnarray}
       \mathcal{A} [\psi^*,\psi] &=&
  \int_{0}^{\hbar\beta}d\tau\int d^D x 
     ~\psi^*(\VEC{x},\tau)\nonumber\\
&&   \times\bigg[\hbar\pdif{\tau}{-}\frac{\hbar^2\Delta}{2M}{+}\frac{M}{2}\omega^2\VEC{x}^2{-}\mu     
     \bigg]\psi(\VEC{x},\tau)\,.\label{klass.wirkung.felder.imaginaerzeit}
\end{eqnarray}
We evaluate the functional integral by using the background method \cite{Morette,DeWitt}. To this end we divide the 
fields $\psi^*(\VEC{x},\tau)$, $\psi(\VEC{x},\tau)$ into field expectation values $\Psi^*(\VEC{x},\tau)$, 
$\Psi(\VEC{x},\tau)$, which we identify later on with the macroscopic occupation of the ground state, and fluctuations 
$\delta\psi^*(\VEC{x},\tau)$, $\delta\psi(\VEC{x},\tau)$:
\begin{eqnarray}
  \psi^*(\VEC{x},\tau)&=&\psixtaust+\delta\psi^*(\VEC{x},\tau)\,,\nonumber\\ \psi(\VEC{x},\tau)&=
&\psixtau+\delta\psi(\VEC{x},\tau)\,.\label{background.zerlegung.felder}
\end{eqnarray}
Note that field expectation values and fluctuations have to satisfy the condition \cite{Yukalov/Graham}
\begin{equation}
   \int d^Dx~\Psi^*(\VEC{x},\tau)\delta\psi(\VEC{x},\tau)=0\,.\label{mittelwert.senkrecht.felder}
\end{equation}
Using (\ref{background.zerlegung.felder}) together with (\ref{grosskan.Z.funtionalint.allg}) we arrive at 
\begin{equation}
    Z=e^{-\mathcal A[\Psi^*,\Psi]/\hbar}\oint\mathcal D\delta\psi^* \mathcal D\delta\psi\,
e^{-\mathcal A[\delta\psi^*,\delta\psi]/\hbar}\,.\label{Z.mit.background}
\end{equation}
Now we decompose the fluctuations $\delta\psi(\VEC{x},\tau)$ into the one-particle eigenstates 
$\psi_\VEC{n}(\VEC{x})$ of the system and apply an additional Matsubara decomposition:
\begin{equation}
    \delta\psi(\VEC{x},\tau)=\sum_{\VEC{n}\neq 0}
    \sum_{m=-\infty}^{\infty}c_{\VEC{n},m}\psi_{\VEC{n}}
    (\VEC{x}) \frac{e^{-i\omega_m \tau}}{\sqrt{\hbar\beta}} \label{feld.separation}
\end{equation}
with the Matsubara frequencies $\omega_m{=}2\pi m/\hbar\beta$.
Note that we explicitly do not sum   over the ground state of the system in (\ref{feld.separation}), 
as we have to satisfy 
condition (\ref{mittelwert.senkrecht.felder}). With this, the measure of the functional integration 
(\ref{Z.mit.background}) turns into
\begin{equation}
   \oint\mathcal D\delta\psi^*\mathcal D\delta\psi=
   \prod_{\VEC{n}\neq 0}\prod_{m=-\infty}^{\infty}\int \frac{d c_{\VEC{n},m}^* d c_{\VEC{n},m}}{2\pi \hbar \beta}\,.
\label{Eq.14}
\end{equation}
The integration over the expansion coefficients $c_{\VEC{n},m}^*$, $c_{\VEC{n},m}$ in (\ref{Eq.14}) is now Gaussian 
and can be performed. The effective action is then found by applying the logarithm to the partition 
function: $\Gamma[\Psi^*,\Psi]{=}-(\log Z)/\beta$. With this one gets
\begin{eqnarray}
 \Gamma[\Psi^*,\Psi]&{=}&\frac{1}{\hbar\beta}\mathcal A[\Psi^*,\Psi]\label{eff.action}\\
&&{-}\frac{1}{\beta}\sum\limits_{k=1}^{\infty}
	       \frac{e^{\beta(\mu-E_0)k}}{k}\left[ \frac{1}{(1{-}e^{-\hbar\omega\beta
	        k})^D}{-}1\right]   \nonumber\,.
\end{eqnarray}
This effective action yields the grand-canonical potential $\Omega$ if it is evaluated for extremised field expectation 
values: $\Omega{=}\Gamma[\Psi_{\rm e}^*,\Psi_{\rm e}]$.
An extremization of (\ref{eff.action}) with respect to $\Psi^*(\VEC{x},\tau)$ leads to
\begin{equation}
   \left\lbrace \hbar\pdif{\tau}-\frac{\hbar^2\Delta}{2M}+\frac{M}{2}\omega^2\VEC{x}^2-\mu\right\rbrace 
\Psi_{\rm e}(\VEC{x},\tau)=0\,.
     \label{extremalisierungs.bed.fuer.mittelwert.allg}
\end{equation}
This equation has the eigenstates $\psi_\VEC{n}(\VEC{x})$ with $\mu{=}E_{\VEC{n}}$ as  non-trivial solutions which 
are periodic in imaginary time. We choose the ground state $\psi_0(\VEC{x}){=}\left(M\omega/\pi\hbar\right)^{D/4}
\exp\left( {-}M\omega\VEC{x}^2/2\hbar\right) $ to be the physically meaningful solution and normalise 
$\Psi_{\rm e}(\VEC{x},\tau)$  to the number of atoms in the ground state $N_0$:
\begin{equation}
 \Psi_{\rm e}(\VEC{x},\tau){=}\sqrt{N_0}\psi_0(\VEC{x})\label{Psi_e}\,.
\end{equation}
As (\ref{extremalisierungs.bed.fuer.mittelwert.allg}) and (\ref{Psi_e}) lead to the algebraic equation 
\begin{equation}
 (E_0{-}\mu)\sqrt{N_0}{=}0\,, 
\end{equation}
we obtain two different phases. In the gas phase we have $N_0{=}0$, whereas in the superfluid phase with $N_0{\neq}0$ 
the chemical potential $\mu$ must be equal to the ground-state energy $E_0$.
The critical temperature $T_c$ occurs at the borderline between both phases, so it follows from the particle
number equation by setting both $N_0=0$ and $\mu=E_0$. To this end
we combine (\ref{eff.action}) and (\ref{Psi_e}) and get for the grand-canonical potential 
\begin{eqnarray}
 \Omega&{=}&(E_0-\mu)N_0\nonumber\\
&&{-}\frac{1}{\beta}\sum\limits_{k=1}^{\infty}
	       \frac{e^{\beta(\mu-E_0)k}}{k}\left[ \frac{1}{(1{-}e^{-\hbar\omega\beta
	        k})^D}{-}1\right]   \,.\label{Omega_bg}
\end{eqnarray}
The first term represents the contribution of the macroscopically occupied ground state of the system, whereas the 
second term describes the thermal contributions of all excited states. This should be compared with 
(\ref{freie.energie.harm}) where the ground state is treated like all other states.
\section{Thermodynamics}
It is now possible to derive all thermodynamic quantities from (\ref{Omega_bg}) within this framework.
The particle number equation $N{=}{-}\partial \Omega/\partial\mu$ reads
\begin{equation}
   N=
      N_0+\sum\limits_{k=1}^{\infty}e^{\beta(\mu-E_0)k}\left[
	       \frac{1}{(1-e^{-\hbar\omega\beta k})^D}{-}1\right]\,. \label{korr.teilchenzahlgl.mit.op}
\end{equation}
The internal energy  follows from the Legendre transformation $U{=}\Omega{+}TS{+}\mu N$:
\begin{eqnarray}
&&  U=E_0N_0{+}D\hbar\omega\sum\limits_{k=1}^{\infty}e^{\beta(\mu{-}E_0)k}
  \bigg\{
  \frac{1}{2}\left[ \frac{1}{(1{-}e^{-\hbar\omega\beta k})^D}{-}1\right]\nonumber\\
&&  \phantom{U=}{+}\frac{e^{-\hbar\omega\beta k}}{(1{-}e^{-\hbar\omega\beta k})^{D+1}}
  \bigg\} 
  \,.\label{korr.innere.energie.harm}
\end{eqnarray}
Both the particle number $N$ and the internal energy $U$ can be expressed by the auxiliary functions
\begin{equation}
   I(A,b,D)=\sum_{k=1}^\infty \frac{ e^{-bAk}}{(1-e^{-bk})^D}\,.
   \label{anh:I.allg}
\end{equation}
The particle number (\ref{korr.teilchenzahlgl.mit.op}) turns into
\begin{equation}
   N=N_0+\sum\limits_{l=1}^{D}(-1)^{l+1}\binom D {l}I(l-\mu',b,D)\,,\label{N.mit.I}
\end{equation}
where we have used the abbreviations $b{=}\hbar\omega\beta$ and $\mu'{=}(\mu{-}E_0)/\hbar\omega$.
For the internal energy (\ref{korr.innere.energie.harm}) we get correspondingly
\begin{eqnarray}
  U&=&E_0N_0+D\hbar\omega\bigg[\frac{1}{2}\sum\limits_{l=1}^{D}(-1)^{l+1}\binom D {l}I(l{-}\mu',b,D)\nonumber\\
 &&    +I(1{-}\mu',b,D{+}1)\bigg] \,.\label{U.mit.I}
\end{eqnarray}
With this it is possible to calculate useful analytic approximations for the critical temperature $T_c$, as well as the 
temperature dependence of both the condensate fraction $N_0/N$  and the heat capacity $C_V$ for different 
numbers of spatial 
dimensions $D$. To this end we assume that the difference $\hbar\omega$ between different energy levels of the 
harmonic oscillator is small compared to the average thermal energy $k_BT$. This yields the semiclassical 
condition $0{<} b{=}\hbar\omega\beta{\ll}1$, which is well fulfilled for present-day experiments.

In order to describe the superfluid phase we apply the limit $\mu{\uparrow}E_0$ to the particle number equation 
(\ref{N.mit.I}) for $D{=}1,2,3$ and use the formulas (\ref{I1})--(\ref{I3}) which are derived in the Appendix 
within the semiclassical approximation.  Furthermore, we expand the polylogarithms for small $b{=}\hbar\omega\beta$ 
by using the Robinson formula \cite{Robinson}
\begin{eqnarray}
\zeta_l ( e^{-a}) {=} \frac{(-a)^{l-1}}{(l{-}1)!}
\bigg\{
\sum_{k=1}^{l-1} \frac{1}{k} {-}  \ln a \bigg\}
{+} \hspace*{-3mm}
\sum_{\begin{array}{c} \mbox{}\\[-5mm]  {\scriptstyle k=0}
\\[-2mm]  {\scriptstyle k\neq l-1} \end{array}}^\infty
\hspace*{-4mm} \frac{(-a)^k}{k!}\zeta (l{-}k)\label{Robinson}\,.
\end{eqnarray}
 Thus, we get for $T{\leq} T_c$
\begin{eqnarray}
 N\big\vert_{D=1}
  &=&N_0{+}  \frac{\gamma{-}\ln(\hbar\omega\beta)}{\hbar\omega\beta}
          {+}\dots
	  \,,\label{teilchenzahl.harm.mit.op.korr.berechnet.t<tc,d=1}
\end{eqnarray}
\begin{eqnarray}
 N\big\vert_{D=2}
  &=&N_0{+}  \frac{\zeta(2)}{(\hbar\omega\beta)^2}
          {+}\frac{ {-}\ln(\hbar\omega\beta){+}\gamma{-}1/2}{\hbar\omega\beta}{+}\dots
	  \,,\qquad\label{teilchenzahl.harm.mit.op.korr.berechnet.t<tc,d=2}
\end{eqnarray}
\begin{eqnarray}
 N\big\vert_{D=3}
  &=&N_0+  \frac{\zeta(3)}{(\hbar\omega\beta)^3}
          {+}\frac{3\zeta(2)}{2(\hbar\omega\beta)^2}\nonumber\\
   &&       {+}\frac{ {-}\ln(\hbar\omega\beta){+}\gamma{-}19/24}{\hbar\omega\beta}{+}\dots
	  \,.\qquad\label{teilchenzahl.harm.mit.op.korr.berechnet.t<tc}
\end{eqnarray}
Here $\zeta(z)$ denotes the Riemann zeta function and $\gamma{=}0.5772{\dots}$ is Euler's constant.
Surprisingly, our results 
(\ref{teilchenzahl.harm.mit.op.korr.berechnet.t<tc,d=1})--(\ref{teilchenzahl.harm.mit.op.korr.berechnet.t<tc}) 
coincide with the findings of two different approaches which are reviewed and compared in Ref.~\cite{Holthaus3}.
The first one is a master equation approach to canonical condensate statistics which is based on an analogy
to the laser phase transtion \cite{Scully1,Scully2}. It yields accurate results even for small systems, is valid
for all temperature, but is partly numerical. The second one, which is entirely based on considering the particle-number
distribution, is fully analytical, but is limited to temperatures below $T_c$ \cite{Holthaus2,Holthaus5}.
\subsection{Critical Temperature}\label{sec:D=3}
Setting $N_0{=}0$ in 
(\ref{teilchenzahl.harm.mit.op.korr.berechnet.t<tc,d=1})--(\ref{teilchenzahl.harm.mit.op.korr.berechnet.t<tc}), 
one obtains for the critical temperature 
\begin{equation}
  T_c\big\vert_{D=1}=\frac{\hbar\omega}{k_B}\frac{N}{\gamma{-}\ln(\hbar\omega/k_BT_c)}{+}\dots\,,\label{tc.d=1}
\end{equation}
\begin{eqnarray}
   T_c\big\vert_{D=2}&=&\frac{\hbar\omega}{k_B}\left( \frac{N}{\zeta(2)}\right)^{\frac{1}{2}}
             \bigg\{ 1{-}\left( \frac{\zeta(2)}{N}\right)^{\frac{1}{2}}\frac{1}{2\zeta(2)}
        \nonumber\\       
     &&\times\bigg[{-}\frac{1}{2}\ln\left(\frac{\zeta(2)}{N}\right){+}\gamma{-}\frac{1}{2}  
      \bigg] 
        \bigg\}+\dots\,,\label{kritische.temp.harm.mit.op.semikl,d=2}
\end{eqnarray}
\begin{eqnarray}
   &&T_c\big\vert_{D=3}=\frac{\hbar\omega}{k_B}\left( \frac{N}{\zeta(3)}\right)^{\frac{1}{3}}
             \bigg\{ 1{-}\left( \frac{\zeta(3)}{N}\right)^{\frac{1}{3}}\frac{\zeta(2)}{2\zeta(3)}    
         \nonumber\\
   &&  {-}\left( \frac{\zeta(3)}{N}\right)^{\frac{2}{3}}   
\frac{1}{3\zeta(3)}
       \left[{-}\frac{1}{3}\ln\left(\frac{\zeta(3)}{N} \right){+}\gamma{-}\frac{19}{24}  
       {-}\frac{3\zeta(2)^2}{4\zeta(3)}\right] 
        \bigg\}\nonumber\\
&&+\dots\,.\label{kritische.temp.harm.mit.op.semikl}
\end{eqnarray}
\begin{figure}[t]
 \begin{minipage}{0.7\linewidth}
\setlength{\unitlength}{1\linewidth}
\begin{picture}(1,0.7)
\includegraphics[width=1\linewidth]{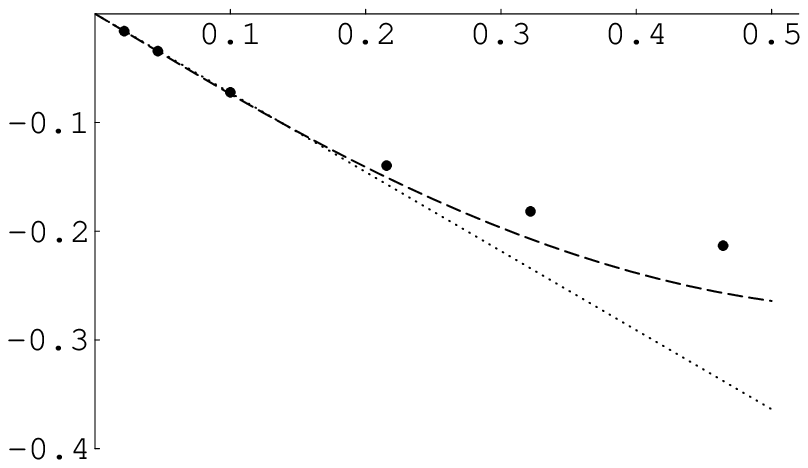}
 \put(-1.05,-0.03){$(T_c{-}T_c^{(0)})/T_c^{(0)}$}
   \put(0.01,0.57){$N^{-1/3}$}
\end{picture}
\end{minipage}
\caption{Finite-size corrections $(T_c{-}T_c^{(0)})/T_c^{(0)}$ taken from Eq.~(\ref{kritische.temp.harm.mit.op.semikl})  
versus $N^{-1/3}$ for $D{=}3$ dimensions. The dotted (dashed) line includes the first (the first two) finite-size 
correction(s). The bullets correspond to the critical temperature which is  numerically determined from 
(\ref{N.mit.geometrischer.reihe.harm}) and (\ref{cond.fraction}) of the theory without order parameter. 
}
\label{fs-cor}
\end{figure}

\noindent In $D{=}1$ dimension  the critical temperature $T_c$ follows from numerically solving   the implicit 
Eq.~(\ref{tc.d=1}). Higher corrections to (\ref{tc.d=1}) are of the order $\mathcal O(N^{-1})$, so they are  small 
and  can be neglected.
Note that our result (\ref{tc.d=1}) differs slightly from the corresponding finding of Refs.~\cite{Ketterle,Yukalov}.
For $D{=}2,3$ one defines the thermodynamic limit $N{\rightarrow}\infty$ in such a way that the leading order of 
(\ref{kritische.temp.harm.mit.op.semikl,d=2}) and (\ref{kritische.temp.harm.mit.op.semikl}), i.~e. 
$T_c^{(0)}{=}\hbar\omega N^{1/D}/k_B\zeta(3)^{1/D}$, remains constant. Thus, when the particle number $N$ is sent 
to infinity  the trap frequency $\omega$ has to approach zero in such a way that the product $N\omega^{D}$ is kept 
fixed.
In this case the quantity $b_c^{(0)}{=}\hbar\omega\beta_c^{(0)}$ is, indeed, small as has been assumed above.
Higher orders of  (\ref{kritische.temp.harm.mit.op.semikl,d=2}) and (\ref{kritische.temp.harm.mit.op.semikl})  are 
called finite-size corrections.  We note that additional nontrivial logarithmic dependences on the particle number $N$ 
occur which do not follow from the standard semiclassical approximation \cite{Pitaevskii,Pethick,Ketterle,Grossmann}. In 
Fig.~\ref{fs-cor} the critical temperature in $D{=}3$ dimensions is plotted up to the first and the second order for 
different particle numbers $N$ and compared with the corresponding finite-size corrections which are obtained 
numerically from the theory without order parameter from (\ref{N.mit.geometrischer.reihe.harm}) and 
(\ref{cond.fraction}). Combining the first and second finite-size corrections from 
(\ref{kritische.temp.harm.mit.op.semikl}) yields a better agreement with the theory without order parameter than the 
first finite-size correction alone.
Furthermore, we read off from Fig.~\ref{fs-cor} that, for particle numbers larger than about $N{=}10^3$, the analytic 
formula (\ref{kritische.temp.harm.mit.op.semikl}) yields values for the critical temperature which coincide with the 
corresponding results of the theory without order parameter for all practical purposes. Thus, although introducing an 
order parameter for studying finite-size effects represents an essential approximation, its findings do not differ from 
the results of the theory without an order parameter for experimentally realistic system sizes.
Note that the first correction of (\ref{kritische.temp.harm.mit.op.semikl}) was already found some time ago 
\cite{Grossmann,Ketterle}, whereas the second correction has only recently been found  \cite[Appendix 7A]{Kleinert}.
\subsection{Condensate Fraction}
The new feature of our approach is that it is  applicable in the whole temperature regime.
For instance, one gets for the condensate fraction $N_0/N$ from 
(\ref{teilchenzahl.harm.mit.op.korr.berechnet.t<tc,d=1})--(\ref{kritische.temp.harm.mit.op.semikl})
\begin{equation}
\frac{N_0}{N}\big\vert_{D=1}=1{-}\frac{T}{T_c}\left[
1{-}\frac{\ln(T_c/T)}{\gamma{-}\ln(\hbar\omega\beta_c)}
\right]  {+}\dots
\,,\label{kondensat.harm.mit.op.bezogen.auf.tc.D=1}
\end{equation}
\begin{eqnarray}
  &&\frac{N_0}{N}\big\vert_{D=2}=
               1{-}\left( \frac{T}{T_c}\right)^2
                {-}\left(\frac{\zeta(2)}{N} \right)^{\frac{1}{2}}\bigg\{ \bigg[\frac{T}{T_c}{-}\left( 
               \frac{T}{T_c}\right)^2  \bigg]\nonumber\\
         &&      \times\frac{-\ln(\zeta(2)/N)/2+\gamma-1/2 }{\zeta(2)}{+}
\frac{T}{T_c}\frac{\ln(T/T_c)}{\zeta(2)}\bigg\}{+}\dots\,,\qquad \label{kondensat.harm.mit.op.bezogen.auf.tc.D=2}
\end{eqnarray}
\begin{eqnarray}
  &&\frac{N_0}{N}\big\vert_{D=3}=
               1{-}\left( \frac{T}{T_c}\right)^3{-}  \left( 
	       \frac{\zeta(3)}{N}\right)^{\frac{1}{3}}\frac{3}{2}\frac{\zeta(2)}{\zeta(3)}
	       \bigg[\left(\frac{T}{T_c} \right)^2 \nonumber\\ 
	&&    {-} \left( \frac{T}{T_c}\right)^3  
	       \bigg]{-}\left( \frac{\zeta(3)}{N}\right)^{\frac{2}{3}}\frac{1}{\zeta(3)}
	       \bigg\{ 
                {-}\frac{3}{2}\frac{\zeta(2)^2}{\zeta(3)}
	       \bigg[\left(\frac{T}{T_c} \right)^2  \nonumber \\
        &&   {-}\left( \frac{T}{T_c}\right)^3 
	       \bigg] 
                  {+}\bigg[\frac{T}{T_c} {-}\left( \frac{T}{T_c}\right)^3  
	       \bigg]\left[{-}\frac{1}{3}\ln\left( 
	       \frac{\zeta(3)}{N}\right){+}\gamma{-}\frac{19}{24} \right]  \nonumber\\
	&&      {+}\frac{T}{T_c}\ln\left( \frac{T}{T_c} \right)	      
	       \bigg\}{+}\dots \,.\label{kondensat.harm.mit.op.bezogen.auf.tc}
\end{eqnarray}
\begin{figure}[t]
\begin{center}
\begin{minipage}[b]{0.45\linewidth}
\setlength{\unitlength}{1\linewidth}
\begin{picture}(1,0.6)
\put(0,0.63){\tiny$N_0/N$}
  \put(0.9,-0.05){\tiny$ T[{\rm \mu K}]$}
\includegraphics[width=1\linewidth]{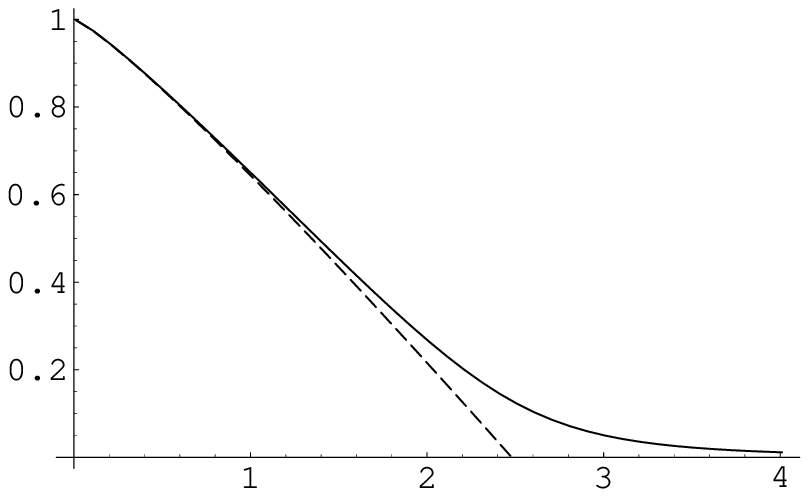}
        \put(-0.8,0.15){\small $\rm a)$}
\end{picture}
\end{minipage}
\hspace{0.00\linewidth}
\begin{minipage}[b]{0.45\linewidth}
\setlength{\unitlength}{1\linewidth}
\begin{picture}(1,0.6)
\put(-0.03,0.63){\tiny$N_0/N$}
  \put(0.9,-0.05){\tiny$ T[{\rm nK}]$}
\includegraphics[width=1\linewidth]{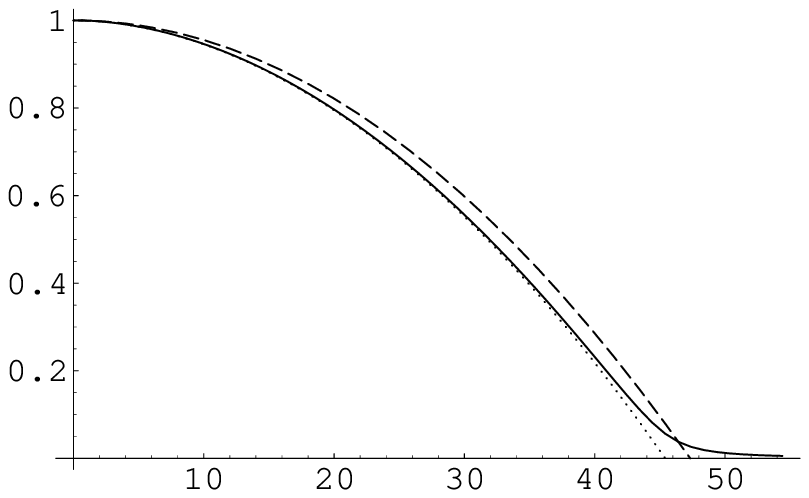}
         \put(-0.8,0.15){\small $\rm b)$}
\end{picture}
\end{minipage}
\end{center}
\caption{Condensate fraction $N_0/N$ versus temperature $T$ for a) $D{=}1$ dimension and $N{=}10^4$ and for b) $D{=}2$  
dimensions and $N{=}10^3$ for $\omega{=}2\pi{\cdot}40~{\rm Hz}$. The solid lines describe Eq.~(\ref{cond.fraction}) 
from the theory without order parameter, the dashed lines correspond to the leading orders of 
(\ref{kondensat.harm.mit.op.bezogen.auf.tc.D=1}), (\ref{kondensat.harm.mit.op.bezogen.auf.tc.D=2}) and the 
dotted line also includes the first finite-size correction.
}
\label{abb:cond-fraction}
\end{figure}

\noindent Figure \ref{abb:cond-fraction} shows how the approximation  to describe a finite system with an order 
parameter deviates from the original grand-canonical approach without order parameter.
In Fig.~\ref{abb:cond-fraction} a) one observes in $D{=}1$ dimension that the approximation introduced by 
the decomposition 
(\ref{background.zerlegung.felder}) is still noticeable at $T{\approx}T_c$ for $N{=}10^4$ particles. For $D{=}2$ 
dimensions, however, this effect becomes much smaller already for $N{=}10^3$ particles as shown in 
Fig.~\ref{abb:cond-fraction} b). Note that Eq.~(\ref{kondensat.harm.mit.op.bezogen.auf.tc.D=2}) including the first 
finite-size correction yields a better agreement with the theory without order parameter than the already known 
leading order of (\ref{kondensat.harm.mit.op.bezogen.auf.tc.D=2}).
\subsection{Heat Capacity}
\begin{figure}[t]
 \begin{minipage}{0.7\linewidth}
\setlength{\unitlength}{1\linewidth}
\begin{picture}(1,0.7)
\includegraphics[width=1\linewidth]{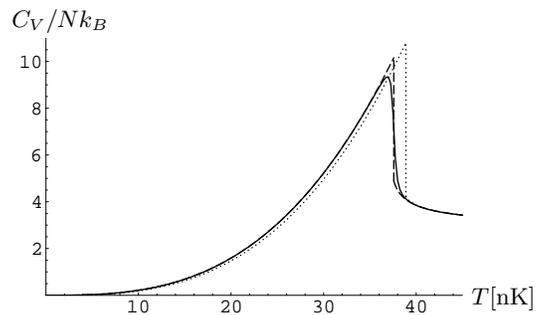}
 \put(-1,0.64){$C_V/Nk_B$}
   \put(0.01,0.03){$T[{\rm nK}]$}
\end{picture}
\end{minipage}
\caption{Heat capacity $C_V/Nk_B$ versus Temperature $T$ in $D{=}3$ dimensions for $\omega{=}2\pi{\cdot}40~{\rm Hz}$ 
and $N{=}10^4$. The dotted (dashed) line is taken from Eqs.~(\ref{Cv.T<Tc}), (\ref{lim.t.von.oben.tc.Cv.fertig.1.ord}) 
and includes the leading order (the first two leading orders). The solid line corresponds to (\ref{Cv.oo.numerisch})  
from the theory without order parameter.}
\label{pub-cv}
\end{figure}

The heat capacity $C_V$ can be found from (\ref{U.mit.I}) by using the relation 
$C_V{=}\partial U/\partial T\big\vert_{V,N}$. One has to take into account two different regimes. For $T{<}T_c$ the 
chemical potential $\mu$ is fixed and the number of atoms in the ground state $N_0$ depends on temperature.  
With this we get 
 \begin{eqnarray}
 &&   C_{V,\,T\leq T_c}\big\vert_{D=1}=Nk_B\frac{2\zeta(2)}{\gamma{-}\ln(\hbar\omega\beta_c)}\frac{T}{T_c}{+}\dots\,,
 \label{Cv.T<Tc,D=1}
\end{eqnarray}
\begin{eqnarray}
 &&   C_{V,\,T\leq T_c}\big\vert_{D=2}=2Nk_B\left(\frac{T}{T_c} \right)^2\bigg\{\frac{3\zeta(3)}{\zeta(2)}
   {-}\left(\frac{\zeta(3)}{N} \right)^{\frac{1}{2}}   \nonumber \\
&&  
    \times\frac{3\zeta(3)}{\zeta(2)^2}\left[
     {-}\frac{1}{2}\ln\left(\frac{\zeta(2)}{N}\right){+}\gamma{-}\frac{1}{2} 
    \right]    
     \bigg\}{+}\dots  
     \,,\qquad
 \label{Cv.T<Tc,D=2}
\end{eqnarray}
\begin{eqnarray}
 &&   C_{V,\,T\leq T_c}\big\vert_{D=3}=3Nk_B \left(\frac{T}{T_c} \right)^3\bigg\lbrace\frac{4\zeta(4)}{\zeta(3)}
   \nonumber \\
&&  {+}\left(\frac{\zeta(3)}{N} \right)^{\frac{1}{3}} 
    \bigg[ \frac{3T_c}{T} {-}\frac{6\,\zeta(4)\zeta(2)}{\zeta(3)^2} 
                    \bigg]    
     \bigg\rbrace{+}\dots  
     \,.
 \label{Cv.T<Tc}
\end{eqnarray}
For $T{>}T_c$, on the other hand,  $N_0$ vanishes and $\mu$ depends on temperature. Thus,
the heat capacity for $T{>}T_c$  still depends 
explicitly on $T$ and $\mu$. However, we can analytically work out the limit $T{\downarrow}T_c$ and obtain
\begin{equation}
\lim_{T\downarrow T_c}C_{V,\,T>T_c}\big\vert_{D=1}
{=}Nk_B\frac{2\zeta(2)}{\gamma{-}\ln(\hbar\omega\beta_c)}{+}\dots
 \label{lim.t.von.oben.tc.Cv.fertig.1.ord,D=1}\,,
\end{equation}
\begin{eqnarray}
   &&\lim_{T\downarrow 
   T_c}C_{V,\,T>T_c}\big\vert_{D=2}\nonumber\\
&&{=}2Nk_B\bigg(3\frac{\zeta(3)}{\zeta(2)}
{-}\frac{2\zeta(2)}{1{+}\gamma{+}\zeta(2){-}\ln(\zeta(2)/N)/2}
\nonumber\\
&&{+}\left(\frac{\zeta(2)}{N} \right)^{\frac{1}{2}}\bigg\{
{-}\frac{3\zeta(3)}{\zeta(2)^2}\left[{-}\frac{1}{2}\ln\left(\frac{\zeta(2)}{N} \right){+}\gamma{-}\frac{1}{2}  \right] \nonumber\\
&&{+}\frac{{-}\ln(\zeta(2)/N)/2{+}5/2{+}3\gamma{-}\ln 2{+}2\zeta(2)}{1{+}\gamma{+}\zeta(2){-}\ln(\zeta(2)/N)/2}\nonumber\\
&&{-}\frac{{-}\ln(\zeta(2)/N)/2{+}\gamma{-}1/2{+}\zeta(2)}
{[1{+}\gamma{+}\zeta(2){-}\ln(\zeta(2)/N)/2]^2}\bigg\}\bigg){+}\dots\,, \label{lim.t.von.oben.tc.Cv.fertig.1.ord,D=2}
\end{eqnarray}
\begin{eqnarray}
   &&\lim_{T\downarrow 
   T_c}C_{V,\,T>T_c}\big\vert_{D=3}{=}3Nk_B\bigg(4\frac{\zeta(4)}{\zeta(3)}{-}3\frac{\zeta(3)}{\zeta(2)}
   {+}\left( 
   \frac{\zeta(3)}{N}\right)^{\frac{1}{3}}\nonumber\\
&&\times\bigg\{\frac{3}{2} {-}6\frac{\zeta(4)\zeta(2)}
   {\zeta(3)^2}{+}\frac{3\zeta(3)}{\zeta(2)^2}\bigg[{-}\frac{1}{2}\ln\left(\frac{\zeta(3)}{N} \right) 
     {+}\frac{5}{4}{+}\zeta(2)\nonumber\\
  && {+}\frac{3}{2}\gamma \bigg]\bigg\}\bigg){+}\dots\,. \label{lim.t.von.oben.tc.Cv.fertig.1.ord}
\end{eqnarray}
Thus, the heat capacity has a discontinuity at $T{=}T_c$ in $D{=}2,3$ which follows 
from (\ref{Cv.T<Tc,D=2}), (\ref{Cv.T<Tc}), (\ref{lim.t.von.oben.tc.Cv.fertig.1.ord,D=2}), and 
(\ref{lim.t.von.oben.tc.Cv.fertig.1.ord}):
\begin{eqnarray}
  &&\Delta C_{V}\big\vert_{D=2}
  {=} 2Nk_B\bigg\{\frac{2\zeta(2)}{1{+}\gamma{+}\zeta(2){-}\ln(\zeta(2)/N)/2}\nonumber\\
&&  {-} \left(\frac{\zeta(2)}{N} \right)^{\frac{1}{2}}\bigg[
{-}\frac{{-}\ln(\zeta(2)/N)/2{+}\gamma{-}1/2{+}\zeta(2)}{[1{+}\gamma{+}\zeta(2){-}\ln(\zeta(2)/N)/2]^2}
  \nonumber\\
&&
{+}\frac{{-}\ln(\zeta(2)/N)/2{+}5{/}2{+}3\gamma{-}\ln 2{+}2\zeta(2)}{1{+}\gamma{+}\zeta(2){-}\ln(\zeta(2)/N)/2}
\bigg]\bigg\}{+}\dots\,,\label{Cv.gap.fertig,D=2}\qquad
\end{eqnarray}
 \begin{eqnarray}
  &&\Delta C_{V}\big\vert_{D=3}
  {=} 3Nk_B\bigg\{\frac{3\zeta(3)}{\zeta(2)}{-}\left( \frac{\zeta(3)}{N}\right)^{\frac{1}{3}}
  \frac{3\zeta(3)}{\zeta(2)^2}
  \nonumber\\
&& \times\bigg[ {-}\frac{1}{2}\ln\left( 
  \frac{\zeta(3)}{N}\right){+}\frac{5}{4}{+}\zeta(2){+}\frac{3}{2}\gamma{-}\frac{\zeta(2)^2}{2\zeta(3)} \bigg] 
   \bigg\}{+}\dots\,.\label{Cv.gap.fertig}\qquad
\end{eqnarray}
In $D{=}3$ dimensions the discontinuity  $\Delta C_V$ remains finite in the thermodynamic limit $N{\rightarrow}\infty$ 
and gets smaller for finite systems. This can be seen in Fig.~\ref{pub-cv}, where the heat capacity taken 
from (\ref{Cv.T<Tc}) and (\ref{lim.t.von.oben.tc.Cv.fertig.1.ord}) including the first finite-size correction are 
compared with the thermodynamic limit. In $D{=}2$ dimensions one has a discontinuity for finite systems which vanishes 
in the thermodynamic limit $N{\rightarrow}\infty$.
Finally, in $D{=}1$ dimension there is no discontinuity $\Delta C_V$ at $T{=}T_c$ at all in leading order in agreement 
with the findings of Ref.~\cite{Yukalov}.  Note that the leading contribution in the heat capacity discontinuities 
$\Delta C_V$  differs slightly from the findings in Ref.~\cite{Yukalov}.

Note that Eqs.~(\ref{Cv.T<Tc,D=1})--(\ref{Cv.T<Tc}) also provide systematic semiclassical expansions for both 
the internal energy $U$ and the entropy $S$ by using the thermodynamic relations 
$C_V{=}\partial U/\partial T\vert_{V,N}{=}T\partial S/\partial T\vert_{V,N}$ \cite{schwabl}. However, one 
does not need to work out the limit $T{\rightarrow}T_c$ separately above and below the critical point as 
both the internal energy $U$ and the entropy $S$ are continuous at $T{=}T_c$.
\section{Conclusions}
In this paper we have extended the usual semiclassical expansion 
\cite{Grossmann,Ketterle,Pitaevskii,Pethick}
for harmonically confined ideal Bose gases. With this we have derived orders of the semiclassical expansion which have 
not yet been accessible using standard semiclassical approaches. This has been shown exemplarily for the critical 
temperature $T_c$ as well as the temperature dependence of the condensate fraction $N_0/N$ and the heat capacity 
$C_V$ in $D{=}1,2,3$ dimensions.  It would be straight-forward to generalize
our findings to anisotropic harmonic trapping potentials, which are used in many experiments
to study Bose-Einstein condensation in $D=2,3$. 
Finally, we note that it would be quite 
instructive to clarify the connection between our field-theoretic approach towards a systematic semiclassical
expansion and the statistical approach of Refs.~\cite{Holthaus1,Holthaus2,Holthaus3,Holthaus4,Holthaus5}
as both seem to be related (see the remark below Eq.~(\ref{teilchenzahl.harm.mit.op.korr.berechnet.t<tc})).
However, this relation is by no means obvious as both the grand-canonical and the canonical approach 
rely on certain approximations.
\section*{Acknowledgement}
We thank Barry Bradlyn,
Konstantin Glaum, Robert Graham, Hagen Kleinert, Walja Korolevski, and Aristeu Lima for useful discussions.
This work has been supported by the  German Research Foundation (DFG) within the Collaborative Research 
Center SFB/TR 12 {\it Symmetries and Universality in Mesoscopic Systems}.
\begin{appendix}
\section{Semiclassical Approximation}
\label{app:Semicl_Approx}
Some thermodynamic properties of the ideal Bose gas in a harmonic trap are expressable in terms of the series
 (\ref{anh:I.allg}).
Generalizing an approach of Ref.~\cite[Appendix 7A]{Kleinert} we work out a systematic semiclassical approximation of 
(\ref{anh:I.allg}) which is valid for $0{<}b{\ll}1$.
\subsection{Euler-MacLaurin Formula}
In the limit of small $b$ it is suggestive to approximate (\ref{anh:I.allg}) by the Euler-MacLaurin formula 
\begin{equation}
    \sum_{k=0}^g f(k)=\int_0^g dk~f(k)+\frac{f(0)+f(g)}{2}+\ldots\label{anh:euler-maclaurin}\,.
\end{equation}
However, this
is not directly possible, as
the integral would diverge if the series starts at $k{=}0$. One way to avoid this divergency is to subtract all 
divergent terms in  (\ref{anh:I.allg}) before replacing the series by an integral.
To this end we expand the denominator of (\ref{anh:I.allg}) for small $b$ 
\begin{equation}
  \frac{1}{(1-e^{-bk})^D}=\sum_{l=0}^{D-1}\frac{C_l(D)}{(bk)^{D-l}}+\mathcal O (b^0)
  \label{anh:nenner.entwickelt}\,,
\end{equation}
where $C_0(D){=}1$, $C_1(D){=}D/2$, $C_2(D){=}D(3D{-}1)/24$, $C_3(D){=}D^2(D{-}1)/48$, $\dots$
are the respective expansion coefficients.
Afterwards we subtract this expansion from (\ref{anh:I.allg}) and add it again:
\begin{eqnarray}
   I(A,b,D)&=&\sum_{k=0}^\infty e^{-bAk}\left[\frac{1}{(1-e^{-bk})^D}
   -\sum_{l=0}^{D-1}\frac{C_l(D)}{(bk)^{D-l}}
   \right]\nonumber\\
&&+\sum_{k=1}^\infty\sum_{l=0}^{D-1}\frac{C_l(D)~e^{-bAk}}{(bk)^{D-l}}+\mathcal O(b^0)\,.
   \label{anh:I.pl.reihe.mi.reihe.k=0}
\end{eqnarray}
As we have extended the summation in the first line from $k{=}1$ to $k{=}0$, we have obtained an error of the order 
$\mathcal O(b^0)$.
Finally, we apply  (\ref{anh:euler-maclaurin}) to the first line in (\ref{anh:I.pl.reihe.mi.reihe.k=0}), where higher 
corrections  can be neglected as they are as well of the order $\mathcal O(b^0)$.
With this we arrive at 
\begin{eqnarray}
   I(A,b,D)&=&\int_0^\infty dk~e^{-bAk}\left[\frac{1}{(1{-}e^{-bk})^D}
   {-}\sum_{l=0}^{D-1}\frac{C_l(D)}{(bk)^{D-l}}
   \right]\nonumber\\
&&+\sum_{k=1}^\infty\sum_{l=0}^{D-1}\frac{C_l(D)~e^{-bAk}}{(bk)^{D-l}}+\mathcal O(b^0)\,.
   \label{anh:I+reihe-reihe.int}
\end{eqnarray}
Both terms in (\ref{anh:I+reihe-reihe.int}) are treated as follows:
\begin{itemize}
\item
We substitute  $x{=}e^{-bk}$ in the first integral of (\ref{anh:I+reihe-reihe.int}). One immediately observes that the  
contribution of this integral is of the order $\mathcal O(b^{-1})$.
Moreover one can reexpress the first integral in terms of Gamma functions by using the Beta function  
\cite[(8.380)]{Gradshteyn}
\begin{equation}
   B(x,y)=\int_0^1 dt\, t^{x-1}(1-t)^{y-1}=\frac{\Gamma(x)\Gamma(y)}{\Gamma(x+y)}\label{anh:betafunktion}\,.
\end{equation}
The remaining integrals in (\ref{anh:I+reihe-reihe.int}) can directly be evaluated using the integral representation 
of the Gamma function \cite[(8.310)]{Gradshteyn}:
\begin{equation}
   \int_0^\infty dk\, k^{x-1}e^{-k a}=\frac{\Gamma(x)}{a^x}
   \label{anh:intdarstellung.gammafunktion.umgeformt}\,.
\end{equation}
\item
The series in the second line of (\ref{anh:I+reihe-reihe.int}) can be expressed in terms of the polylogarithms 
(\ref{polylog}).
\end{itemize}
With this we obtain eventually
\begin{eqnarray}
 &&I(A,b,D){=}\frac{1}{b}\bigg[\frac{\Gamma(1{-}D)\Gamma(A)}{\Gamma(1{+}A{-}D)}    
{-}\sum_{l=0}^{D{-}1}C_l(D)~\Gamma(1{+}l{-}D)\nonumber\\
 &&\times A^{D{-}l{-}1}
   \bigg] {+}\sum_{l=0}^{D-1}\frac{C_l(D)}{b^{D{-}l}}\zeta_{D{-}l}\left(e^{-Ab}\right)
     {+}\mathcal O(b^0)\,.\label{anh:I_0.allg.D}
\end{eqnarray}
\subsection{Dimensional Regularisation}
As the Gamma functions in (\ref{anh:I_0.allg.D}) are divergent for integer dimension $D$, we
apply dimensional regularization \cite{Schulte,Zinn-Justin}.
To this end we set  $D{=}d{-}\epsilon$ with $d{=}1,2,3,4\dots$ and 
consider the limit  $\epsilon{\rightarrow}0$. This leads to
\begin{eqnarray}
   I(A,b,d)&{=}&\lim_{\epsilon\rightarrow 0}\frac{1}{b}\bigg[\frac{\Gamma(1{-}d{+}\epsilon)\Gamma(A)}{\Gamma(1{+}A{-}d
+\epsilon)}\nonumber\\
  && {-}\sum_{l=0}^{d-1}C_l(d{-}\epsilon)~\Gamma(1{+}l{-}d{+}\epsilon)A^{d{-}l{-}1{-}\epsilon}
   \bigg]\nonumber\\
 &&  {+}\sum_{l=0}^{d-1}\frac{C_l(d)}{b^{d{-}l}}\zeta_{d-l}\left(e^{-Ab}\right)
     {+}\mathcal O(b^0)\,.
   \label{anh:I_0.allg.d-e}
\end{eqnarray}
In order to evaluate the limit $\epsilon{\rightarrow}0$ we use the following expansion for the Gamma function
\begin{equation}
   \Gamma({-}n{+}\epsilon)=\frac{(-1)^n}{n!}\left\lbrace\frac{1}{\epsilon}{+}\psi_0(n{+}1){+}\mathcal O(\epsilon^1)  
\right\rbrace
   \label{anh:entw.gammafunktion.allg}
\end{equation}
with $n{=}1,2,3,\dots$  and $\epsilon{>}0$ \cite[(8D.24)]{Schulte}. The digamma function $\psi_0(z)$ is defined as
$\psi_0(z){=}\Gamma^\prime(z)/\Gamma(z)$.
 It satisfies the recursion formula \cite[(8D.6)]{Schulte}
$\psi_0(z){=}1/(z{-}1){+}\psi_0(z{-}1)$,
where $\gamma{=}{-}\psi_0(1){=}0.5772\ldots$ denotes Euler's constant. Moreover, we use the approximation  
$x^\epsilon{=}1{+}\epsilon\ln x{+}\dots$ and expand the coefficients
$
C_l(d{-}\epsilon){=}C_l^{(0)}(d){-}C_l^{(1)}(d)\epsilon{+}\mathcal O(\epsilon^2)
$
for small $\epsilon$, where 
$C_l^{(0)}(d){=}C_l(d)$, $C_0^{(1)}(d){=}0$, $	C_1^{(1)}(d){=}1{/}2$, $C_2^{(1)}(d){=}(6d{-}1)/24$,  
 $C_3^{(1)}(d){=}d(3d{-}2)/48$, $\dots$ are  the respective expansion coefficients.
With this (\ref{anh:I_0.allg.d-e}) reduces to
\begin{eqnarray}
 &&  I(A,b,d){=}\lim_{\epsilon\rightarrow 0}\frac{(-1)^{d-1}}{b}\bigg\{
                \frac{1}{(d{-}1)!}\left[ \prod_{l=1}^{d-1}(A-l)\right] \nonumber\\
	 &&      \times\left[\frac{1}{\epsilon}{-}\psi_0(1{+}A{-}d)+\psi_0(d) \right]
		{-}\sum_{l=0}^{d-1}\frac{(-1)^l C_l^{(0)}(d)}{(d{-}l{-}1)!}\nonumber\\
	&&	\times\bigg[\frac{1}{\epsilon}{+}\psi_0(d-l)
               -\frac{C_l^{(1)}(d)}{C_l^{(0)}(d)}{-}
		\ln(A) \bigg] A^{d-l-1}\bigg\}\nonumber\\
	&&      {+}\sum_{l=0}^{d-1}\frac{C_l^{(0)}(d)}{b^{d{-}l}}\zeta_{d-l}
	\left(e^{-Ab}\right)
     {+}\mathcal O(b^0)\,.
     \label{anh:I0.entwickl.eingesetzt.aber.noch.mit.lim}
\end{eqnarray}
Equation (\ref{anh:I0.entwickl.eingesetzt.aber.noch.mit.lim}) can be further simplified if one uses the identity
\begin{equation}
    \frac{1}{(d{-}1)!}\prod_{l=1}^{d-1}(A-l)=\sum_{l=0}^{d-1}\frac{(-1)^l 
C_l^{(0)}(d)}{(d{-}l{-}1)!}A^{d-l-1}\label{anh:Id.fuer.regular}\,,
\end{equation}
which can be proven by complete induction.
Thus, the terms in (\ref{anh:I0.entwickl.eingesetzt.aber.noch.mit.lim}), which are proportional to 
$1/\epsilon$, cancel and
we obtain a finite result in the limit $\epsilon{\rightarrow}0$. Using the recursion formula 
of the digamma function one arrives at
\begin{eqnarray}
&&   I(A,b,d){=}\frac{1}{b}
               \sum_{l=0}^{d-1}\frac{(-1)^{d-l-1} C_l^{(0)}(d)}{(d{-}l{-}1)!} A^{d-l-1}\nonumber\\
         &&	\times\left[{-}\psi_0(1{+}A{-}d){+}\ln A{+}\frac{C_l^{(1)}(d)}{C_l^{(0)}(d)}
		{+}\sum_{m=d-l}^{d-1}\frac{1}{m} \right]\nonumber\\
	&&	{+}\sum_{l=0}^{d-1}\frac{C_l^{(0)}(d)}{b^{d{-}l}}\zeta_{d-l}\left(e^{-Ab}\right)
     {+}\mathcal O(b^0)\,.
     \label{anh:I0.fertig.allg}
\end{eqnarray}  
Assuming  $0{<}b{\ll}1$ we read off that Eq.~(\ref{anh:I0.fertig.allg}) is, indeed, a good approximation for 
(\ref{anh:I.allg}) as the error is of the order  $\mathcal O(b^0)$.
For different dimensions $d{=}1,2,3,4$ one gets explicitly
\begin{eqnarray}
  &&I(A,b,1)=\frac{1}{b}\left[\ln A{-}\psi_0(A){+}\zeta_1(e^{-Ab}) \right]{+}\mathcal O(b^0)\,,\qquad\label{I1}
\end{eqnarray}
\begin{eqnarray}
  &&I(A,b,2)=\frac{1}{b^2}\zeta_2(e^{-Ab}){+}\frac{1}{b}\bigg\{
              {-}(A{-}1)\left[\ln A{-}\psi_0(A{-}1)
	       \right]\nonumber\\
            &&{+}\frac{3}{2}{+}\zeta_1(e^{-Ab})
               \bigg\} {+}\mathcal O(b^0) \,,\label{I2}
\end{eqnarray}
\begin{eqnarray}
  &&I(A,b,3)=\frac{1}{b^3}\zeta_3(e^{-Ab}){+}\frac{3}{2b^2}\zeta_2(e^{-Ab})  
{+}\frac{1}{b}\bigg\{\frac{1}{2}(A{-}1)\nonumber\\
          &&\times(A{-}2)\left[\ln A{-}\psi_0(A{-}2)
	      \right]  {-}\frac{5}{4}A+\frac{53}{24}+\zeta_1(e^{-Ab})\bigg\}\nonumber\\
&&  {+}\mathcal O(b^0) \,,\label{I3}
\end{eqnarray}
\begin{eqnarray}
  && I(A,b,4)=\frac{1}{b^4}\zeta_4(e^{-Ab}){+}\frac{2}{b^3}\zeta_3(e^{-Ab})
              {+}\frac{11}{6b^2}\zeta_2(e^{-Ab})\nonumber\\
	&&      {+}\frac{1}{b}\bigg\{{+}\frac{7}{12}A^2{-}\frac{179}{72}A{+}\frac{8}{3} {+}\zeta_1(e^{-Ab}) 
	        {-}\frac{1}{6}(A{-}1)(A{-}2) \nonumber\\
	&&  \times(A{-}3)\left[\ln 
	        A{-}\psi_0(A{-}3) \right]
	      \bigg\}  {+}\mathcal O(b^0)\,.\label{I4}
\end{eqnarray}
Note that the respective polylogarithmic functions in (\ref{I1})--(\ref{I4}) have to be evaluated for small $b$
by using the Robinson formula (\ref{Robinson}).
\end{appendix}

\end{document}